\documentclass[a4paper, onecolumn, noarxiv, accepted=2024-04-15]{quantumarticle}
\usepackage[utf8]{inputenc}
\usepackage{color}
\usepackage{bm}
\usepackage{xcolor}
\usepackage{amsmath}
\usepackage{amssymb}
\usepackage{graphicx}
\usepackage[numbers, sort&compress]{natbib}
\usepackage{braket}
\usepackage{psfrag}
\usepackage{tikz}
\usepackage[normalem]{ulem}
\usepackage{qcircuit}
\usepackage{ulem}
\usepackage{multirow}
\usepackage{algorithmic}
\usepackage[ruled,linesnumbered,vlined]{algorithm2e}
\usepackage{hyperref}
\hypersetup{colorlinks=true, citecolor=blue, urlcolor=blue, linkcolor=blue}
\usepackage[caption=false]{subfig}
\usepackage{esint}
\usepackage{kbordermatrix}
\usepackage{tabularx}
\usepackage{booktabs}
\allowdisplaybreaks

\newcommand{\identity}{\mathbb{I}}

\newcommand{\fmax}{f_{\textrm{max}}}
\newcommand{\Usin}{U_{\textrm{sin}}}
\newcommand{\Usqrt}{U_{\textrm{sqrt}}}

\begin{document}

\title{Derivative Pricing using Quantum Signal Processing}

\author{Nikitas Stamatopoulos}
\affiliation{Goldman Sachs, New York, NY}

\author{William J. Zeng}
\affiliation{Goldman Sachs, New York, NY}

\begin{abstract}
	Pricing financial derivatives on quantum computers typically includes quantum arithmetic components which contribute heavily to the quantum resources required by the corresponding circuits.
	In this manuscript, we introduce a method based on Quantum Signal Processing (QSP) to encode financial derivative payoffs directly into quantum amplitudes, alleviating the quantum circuits from the burden of costly quantum arithmetic.
	Compared to current state-of-the-art approaches in the literature, we find that for derivative contracts of practical interest, the application of QSP significantly reduces the required resources across all metrics considered, most notably the total number of T-gates by $\sim 16$x and the number of logical qubits by $\sim 4$x.
	Additionally, we estimate that the logical clock rate needed for quantum advantage is also reduced by a factor of $\sim 5$x.
    Overall, we find that quantum advantage will require $4.7$k logical qubits, and quantum devices that can execute $10^9$ T-gates at a rate of $45$MHz.
	While in this work we focus specifically on the payoff component of the derivative pricing process where the method we present is most readily applicable, similar techniques can be employed to further reduce the resources in other applications, such as state preparation.
\end{abstract}

\maketitle

\section{Introduction}
Financial derivatives can be priced on quantum computers using Quantum Amplitude Estimation (QAE) \cite{brassard2002quantum} with a quadratic speedup compared to classical Monte Carlo methods \cite{montanaro2015quantum,rebentrost2018quantum, Woerner_2019, Stamatopoulos_2020, doriguello_et_al:LIPIcs.TQC.2022.2, Herbert_2022}.
However, the resources required by the resulting quantum circuits have been estimated to be considerably larger than what is expected to be possible with near term devices \cite{chakrabarti2021threshold}.
Possible reduction of these resources has been proposed by replacing quantum arithmetic with lookup tables \cite{qsharpblog}, applying generative methods to prepare relevant probability distributions \cite{Zoufal_2019}, extracting multiple other useful quantities in derivative pricing \cite{Stamatopoulos_2022}, and improving amplitude estimation \cite{suzuki2020amplitude, grinko2021iterative}.

Quantum Signal Processing\footnote{This technique is also referred to as Quantum Singular Value Transformation (QSVT) or Quantum Eigenvalue Transform (QET) depending on the shape/dimensionality of the subsystem transformed. In this manuscript we use QSP as an umbrella term encompassing all transformations of this nature which rely on the same underlying process.} (QSP) is a technique to perform nearly arbitrary polynomial transformations to quantum (sub)systems with low resource overhead \cite{gilyen2019quantum}.
Various quantum algorithms have been unified under the QSP framework \cite{martyn2021grand} and in multiple instances the QSP formulation has led to a significant reduction in resource overheads \cite{Low2017optimal, martyn2023efficient, Lin_2020, Rall_2023, mcardle2022quantum}.
In particular, in Ref.~\cite{mcardle2022quantum} the authors show how to apply QSP for quantum state preparation in cases where the target amplitudes are known functions of a variable whose discretized values have been encoded as computational basis states.
Additionally, the authors demonstrated that this QSP approach significantly outperforms the use of quantum (coherent) arithmetic for the same task in terms of the total quantum resources required, with most pronounced impact on the number of qubits.

Because the quantum oracles involved in derivative pricing inevitably include components heavily relying on quantum arithmetic \cite{chakrabarti2021threshold}, in this manuscript we show how QSP and the ideas from Ref.~\cite{mcardle2022quantum} can be used to reduce the quantum resources needed for derivative pricing for certain very common types of derivatives.
We identify the specific circuit components where QSP can replace quantum arithmetic routines, and provide implementations for all the unitaries involved across the entire applicable domain.
Moreover, we introduce a new unitary to drive the QSP process, leading to further reduction in resources compared to that considered in Ref.~\cite{mcardle2022quantum}.
We examine the impact of this method on concrete derivative applications of practical interest and demonstrate the precise impact on quantum resources by explicitly constructing the corresponding circuits.

The rest of the paper in structured as follows: In Sec.~\ref{sec:quantum_pricing} we describe the core components used for derivative pricing on a quantum computer and highlight the bottlenecks we target in this work.
In Sec.~\ref{sec:main_idea} we outline the QSP framework and present the core ideas used to apply the framework to derivative pricing.
Then, in Sec.~\ref{sec:implementation} we describe the implementation details of the proposed QSP approach and in Sec.~\ref{sec:applications} we demonstrate how to apply the ideas developed to derivative pricing instances of practical interest.
In Sec.~\ref{sec:resources} we calculate the quantum resources needed for quantum advantage in derivative pricing using the method we develop, and contrast them with those of previous approaches in the literature.
Finally, in Sec.~\ref{sec:discussion}, we discuss the implications of this work, highlight its limitations, and outline potential future research directions on the topic.

\section{Derivative Pricing with Quantum Computers}
\label{sec:quantum_pricing}
Financial derivatives are contracts whose value depends on the future performance of one or more underlying assets.
Derivative contracts are typically defined by their \emph{payoff}, which determines the value of the contract for a given future \emph{path} of the underlying assets.
Classically, these contracts are often priced by simulating a large number of possible such paths according to an appropriate stochastic model, computing the payoff for each simulated path, and calculating the expectation value across all future payoffs.
The \emph{fair price} of the derivative contract is then given by that future expectation value, appropriately discounted to take into account the difference between the time the contract is priced and its future value.
For an overview of derivative types, features and classical pricing methodologies, we refer the reader to Ref.~\cite{Hull}.

In a quantum setting, the quantum circuits which drive the pricing process consist of two main components \cite{chakrabarti2021threshold}: 1) loading the probability distribution over paths of the (discretized) stochastic variables involved in the calculation and 2) encoding the payoff of each path into the amplitude of an ancilla qubit.
The first component can be represented by an operator $\mathcal{P}$ which, given stochastic paths $\omega \in \Omega$ where each path occurs with probability $p(\omega)$, it prepares a probability-weighted superposition of all paths

\begin{equation}
	\label{eqn:derivative_path_loading}
	\mathcal{P} :\ket{\vec{0}} \rightarrow \sum_{\omega}\sqrt{p(\omega)}\ket{\omega}.
\end{equation}
The second component is an operator $\mathcal{F}$ which computes the payoff $f(\omega)$ of the derivative on each path $\ket{\omega}$, and encodes that value in the amplitude of an ancilla qubit

\begin{equation}
	\label{eqn:derivative_payoff}
	\mathcal{F}:\ket{\omega}\ket{0} \rightarrow \ket{\omega}\ket{f(\omega)} \left( \sqrt{f(\omega)}\ket{0} + \sqrt{1-f(\omega)}\ket{1} \right).
\end{equation}
The operator $\mathcal{A} = \mathcal{F}\mathcal{P}$

\begin{equation}
	\label{eqn:A_operator}
	\mathcal{A} : \ket{\vec{0}}\ket{0} \rightarrow \left( \sum_{\omega}\sqrt{p(\omega)}\sqrt{f(\omega)}\ket{\omega}\ket{f(\omega)}\ket{0} +  \sum_{\omega}\sqrt{p(\omega)}\sqrt{1-f(\omega)}\ket{\omega}\ket{f(\omega)}\ket{1} \right),
\end{equation}
creates a state such that the probability of measuring $\ket{0}$ in the last qubit is the price of the derivative, represented as the expected value of the payoff, $\mathbb{E}[f] = \sum_{\omega \in \Omega}p(\omega)f(\omega)$.

The implementation of the $\mathcal{F}$ operator of Eq.~\eqref{eqn:derivative_payoff} requires binary quantum arithmetic to compute a digital representation of the payoff $\ket{f(\omega)}$ from a path $\ket{\omega}$, before that value is encoded into the amplitude of the last qubit.
It is this component we will aim to implement using QSP, such that the value $\sqrt{f(\omega)}$ is directly encoded in the amplitude of an ancilla qubit without requiring the computation of $\ket{f(\omega)}$, thus eliminating the costly quantum arithmetic involved.

\section{Quantum Signal Processing}
\label{sec:main_idea}
Given a quantum register $\ket{x}_n$ which stores an $n$-bit representation of a scalar $x \in [0, 1]$, a real function $g: [0,1] \rightarrow [0,1]$ and a unitary which creates the state

\begin{equation}
	\label{eqn:general_amplitude_oracle}
	U\ket{x}_n\ket{0}_{n+1} = \ket{x}_n\left(g(x)\ket{\psi_0}_n\ket{0} + \sqrt{1-g(x)^2}\ket{\psi_1}_n\ket{1} \right),
\end{equation}
where $\ket{\psi_0}_n$ and $\ket{\psi_1}_n$ are normalized quantum states, Quantum Signal Processing (QSP) can be used to apply polynomial transformations to the amplitude $g(x)$.

\begin{figure}[t]
  \centering
  \includegraphics[width=0.85\linewidth]{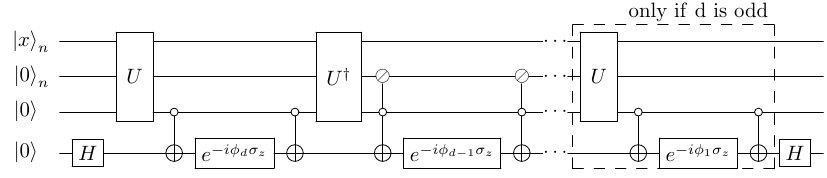}
  \caption{Circuit diagram of the $U^{\vec{\phi}}$ operator of Eq.~\eqref{eqn:qsp_general_transformation} based on the unitary $U$ of Eq.~\eqref{eqn:general_amplitude_oracle}. The choice of angles $\vec{\phi}=(\phi_1, \phi_2, \cdots, \phi_d)$ determines the exact $d$-degree polynomial $P$ applied to the state, such that the probability of measuring $\ket{00}$ in the bottom two qubits if $d$ is odd and $\ket{0}_{n+2}$ in the bottom $n+2$ qubits if $d$ is even, will be given by $|P(g(x))|^2$. We use the $\oslash$ symbol to indicate a unitary that is controlled on every qubit of a quantum register being in the $\ket{0}$ state.}
  \label{fig:qsp_circuit}
\end{figure}

In Ref.~\cite{gilyen2019quantum}, the authors observe that given projectors $\tilde{\Pi} \equiv \identity^{\otimes 2n} \otimes \ket{0}\bra{0}$ and $\Pi \equiv \identity^{\otimes n} \otimes \left(\ket{0}\bra{0} \right)^{\otimes n+1}$, $U$ is a block-encoding of the rank-1 matrix $A = \tilde{\Pi}U\Pi$
\begin{eqnarray}
	\label{eqn:block_encoding}
	U = \kbordermatrix{\mbox{} &\Pi &  \\
	\tilde{\Pi} & A     & \cdot \\
	& \cdot & \cdot
	},
\end{eqnarray}
with a single non-trivial singular value $g(x)$.
They then show that consecutive invocations of $U$ and $U^{\dagger}$, interleaved with projector-controlled rotation operators $\Pi_{\phi}=e^{i\phi(2\Pi - \identity)}$ and $\tilde{\Pi}_{\phi}=e^{i\phi(2\tilde{\Pi} - \identity)}$, can be used to apply polynomial transformations to this singular value.
Specifically, given phase factors $\vec{\phi}=(\phi_1, \phi_2, \cdots, \phi_d)$, define

\begin{align}
	\label{eqn:qsp_general_unitary}
	\mathcal{U}^{\vec{\phi}} = \begin{cases}
					\tilde{\Pi}_{\phi_1}U\displaystyle \prod_{k=1}^{(d-1)/2}\Pi_{\phi_{2k}}U^{\dagger}\tilde{\Pi}_{\phi_{2k+1}}U, & \text{for odd $d$,} \\
					\displaystyle\prod_{k=1}^{d/2}\Pi_{\phi_{2k-1}}U^{\dagger}\tilde{\Pi}_{\phi_{2k}}U, & \text{for even $d$}
	\end{cases}
\end{align}
and
\begin{equation}
	\mathcal{U}_C^{\vec{\phi}} = \left(\mathcal{U}^{\vec{\phi}} \otimes \ket{0}\bra{0} +  \mathcal{U}^{-\vec{\phi}} \otimes \ket{1}\bra{1} \right).
\end{equation}
Then, for any polynomial $P$ satisfying $\textrm{deg}(P) \le d$, $|P(a)| \le 1, \forall a \in [-1, 1]$, and $P$ either even or odd, the authors of Ref.~\cite{gilyen2019quantum} show that there exist phase factors $\vec{\phi}$ \cite{Haah2019product, chao2020finding, dong2021efficient} such that the unitary

\begin{equation}
	\label{eqn:qsp_conditional_unitary}
U^{\vec{\phi}} = (\identity^{\otimes 2n+1} \otimes H)  \mathcal{U}_C^{\vec{\phi}}  (\identity^{\otimes 2n+1} \otimes H),
\end{equation}
applies the polynomial $P$ to the singular values of matrix $A$

\begin{equation}
	\label{eqn:block_encoding_transformation}
U^{\vec{\phi}} =
	\kbordermatrix{\mbox{} & \big(\Pi \otimes \ket{0}\bra{0} \big) &  \\
	\left(\Pi' \otimes \ket{0}\bra{0} \right) & P(A)     & \cdot \\
	& \cdot & \cdot
	}
\end{equation}
where $\Pi'=\tilde{\Pi}$ for odd $d$ and $\Pi'=\Pi$ for even $d$.
Equivalently, $U^{\vec{\phi}}$ creates the state

\begin{equation}
	\label{eqn:qsp_general_transformation}
	U^{\vec{\phi}}\ket{x}_n\ket{0}_{n+1}\ket{0} = \ket{x_n}
	\begin{cases}
			\left(P(g(x))\ket{\psi'_0}_n\ket{0}_2 + \sqrt{1-P(g(x))^2}\ket{\psi'_1}_{n}\ket{0_{\perp}}_2 \right), & \text{for odd $d$,} \\
			& \\
			\left(P(g(x))\ket{0}_{n+2} + \sqrt{1-P(g(x))^2}\ket{0_{\perp}}_{n+2} \right), & \text{for even $d$.}
	\end{cases}
\end{equation}
A circuit description of the operator $U^{\vec{\phi}}$ is shown in Fig.~\ref{fig:qsp_circuit}.

In Ref.~\cite{mcardle2022quantum}, the authors consider the scenario where a target function $f(x)$ is computed by first constructing a unitary of the form of Eq.~\eqref{eqn:general_amplitude_oracle} and then applying a polynomial transformation using QSP approximating the function composition $\left(f \circ g^{-1}\right)(x)$.
Specifically, the unitary in Eq.~\eqref{eqn:general_amplitude_oracle} is taken to be

\begin{equation}
	\label{eqn:sine_amplitude_oracle}
	\Usin\ket{x}_n\ket{0} = \ket{x}_n\left(\sin(x)\ket{0} + \cos(x)\ket{1} \right),
\end{equation}
and QSP transformations based on this unitary are used to prepare states whose amplitudes are functions of $x$, without requiring quantum arithmetic.
With this unitary the polynomial transformation applied with QSP needs to approximate $(f \circ \arcsin)(x)$.
$\Usin$ can be constructed with $n$ controlled-$R$y rotations, and the unitary $\mathcal{U}^{\vec{\phi}}$ of Eq.~\eqref{eqn:qsp_general_unitary} is slightly simpler to implement in this case because the corresponding projectors satisfy $\Pi = \tilde{\Pi} \equiv \identity^{\otimes n} \otimes \ket{0}\bra{0}$.
The choice of $\Usin$ additionally simplifies the final state in Eq.~\eqref{eqn:qsp_general_transformation}, requiring $n$ fewer ancilla qubits, leading to

\begin{equation}
	\label{eqn:qsp_sin_transformation}
	U^{\vec{\phi}}\ket{x}_n\ket{0}_2 = \ket{x_n}\left(P(g(x))\ket{0}_2 + \sqrt{1-P(g(x))^2}\ket{0_{\perp}}_2 \right).
\end{equation}

In this manuscript we instead consider the unitary

\begin{equation}
	\label{eqn:sqrt_amplitude_oracle}
	\Usqrt\ket{x}_n\ket{0}_{n+1} = \ket{x}_n\left(\sqrt{x}\ket{\psi_0}_n\ket{0} + \sqrt{1-x}\ket{\psi_1}_n\ket{1} \right),
\end{equation}
such that QSP needs to approximate the function $f(x^2)$.
The material difference between these two approaches lies mainly in the amount of resources required to construct $\Usin$ and $\Usqrt$, as well as the polynomial degree needed to construct approximations of sufficient accuracy to the target transformation in each case.
Because the details of QSP transformations based on $\Usin$ are analyzed in Ref.~\cite{mcardle2022quantum}, for the rest of this manuscript we focus on the implementation of the QSP transformation driven by $\Usqrt$, and revisit the former when we estimate the resources required to apply this technique in derivative pricing.

In this work we focus on target functions of the form $\left(Ae^x - B\right)/C$ because this particular form is ubiquitous in derivative pricing \cite{Hull}.
This motivation is further illustrated in practice in Sec.~\ref{sec:applications} where we apply these methods to price concrete financial derivatives.
Namely, we are looking for a unitary which prepares the state

\begin{equation}
	\label{eqn:general_exp_state}
	\ket{x}_n\ket{0}_{n+q} \rightarrow \ket{x}_n \left(\sqrt{\frac{Ae^x - B}{C}}\ket{\psi'_0}_n\ket{0}_q + \sqrt{1 - \frac{Ae^x - B}{C}}\ket{\psi'_1}_n\ket{0_{\perp}}_q \right),
\end{equation}
for normalized quantum states $\ket{\psi'_0}_n$, $\ket{\psi'_1}_n$, and $\ket{0_{\perp}}_{q}$ denoting a normalized state orthogonal to $\ket{0}_{q}$.
We assume that constants $A,B,C$ are such that $(Ae^x - B)/{C} \in [0,1]$.

\section{Implementation}
\label{sec:implementation}

In this section we show how to implement the unitary $\Usqrt$ of Eq.~\eqref{eqn:sqrt_amplitude_oracle} and the choice of polynomial $P$ needed in Eq.~\eqref{eqn:qsp_general_transformation} in order to construct the target state Eq.~\eqref{eqn:general_exp_state} under various scenarios.
We assume that the register $\ket{x}_n$ is an $n$-bit, binary, fixed-point representation of a scalar $x$ with $p$ digits to the left of the binary point

\begin{equation}
\label{eqn:fixed_point_repr}
x=\underbrace{x_{n-1}\cdots x_{n-p}}_p.\underbrace{x_{n-p-1} \cdots x_0}_{n-p}.
\end{equation}
Let us first consider the case where $x$ is positive with $p=0$, such that $x<1$

\begin{equation}
	\label{eqn:binary_representation}
	x = \sum_{i=0}^{N-1}\frac{x_i \cdot 2^{i}}{N},
\end{equation}
where $x_i \in {0,1}$ denotes the $i$-th bit of $\ket{x}_n$ and $N=2^n$.
The first step to implement the transformation of Eq.~\eqref{eqn:general_exp_state} is to encode the value $\sqrt{x}$ to the amplitude of an ancilla qubit.
We start with the state $\ket{x}_n$ and add $n$ ancilla qubits in a uniform superposition

\begin{equation}
	\label{eqn:comparator_superposition}
	\ket{x}_n\ket{+}_n = \sum_{j=0}^{N-1}\frac{1}{\sqrt{N}}\ket{x}_n\ket{j}_n.
\end{equation}
We then apply the unitary

\begin{equation}
	\mathcal{C} : \ket{a}\ket{b}\ket{0} \rightarrow \ket{a}\ket{b}\ket{a < b}
\end{equation}
to perform a binary comparison between the two qubit registers of Eq.~\eqref{eqn:comparator_superposition}.
This produces the state

\begin{equation}
	\label{eqn:comparator_expansion}
	\sum_{j=0}^{N-1}\frac{1}{\sqrt{N}}\ket{x}_n\ket{j}_n\ket{N\cdot x < j} = \ket{x}_n \otimes \left[\sum_{j=0}^{N\cdot x-1}\frac{1}{\sqrt{N}}\ket{j}_n\ket{0} + \sum_{j=N\cdot x}^{N-1}\frac{1}{\sqrt{N}}\ket{j}_n\ket{1}  \right],
\end{equation}
with $x$ given by Eq.~\eqref{eqn:binary_representation}.
By inspection, we observe that the probability of measuring $\ket{0}$ in the comparator ancilla qubit is then $\frac{N\cdot x}{N} = x$.
The unitary $\Usqrt = \mathcal{C} (\identity^{\otimes n} \otimes H^{\otimes n} \otimes \identity)$ acting on the state $\ket{x}_n\ket{0}_{n+1}$therefore creates the target state of Eq.~\eqref{eqn:sqrt_amplitude_oracle}

\begin{equation}
	\label{eqn:U_implementation_positive}
	\Usqrt\ket{x}_n\ket{0}_{n+1} = \ket{x}_n\left(\sqrt{x}\ket{\psi_0}_n\ket{0} + \sqrt{1-x}\ket{\psi_1}_n\ket{1} \right),
\end{equation}
where $\ket{\psi_0}_n$ and $\ket{\psi_1}_n$ are normalized quantum states which depend on the value of $x$.
A polynomial $P$ which approximates the function

\begin{equation}
	f(x) = \sqrt{\frac{Ae^{x^2} - B}{C}},
\end{equation}
in the domain $x \in [0,1]$ allows us to approximately prepare the target state given by Eq.~\eqref{eqn:general_exp_state} using Eq.~\eqref{eqn:qsp_general_transformation}.

When $p>0$ such that $x \ge 1$, we can use the same unitary $\Usqrt$ as above, ignoring the location of the binary point $p$, which in this case generates the state

\begin{equation}
	\label{eqn:U_implementation_positive_scaling}
	\Usqrt\ket{x}_n\ket{0}_{n+1} = \ket{x}_n\left(\sqrt{\frac{x}{2^p}}\ket{\psi_0}_n\ket{0} + \sqrt{1-\frac{x}{2^p}}\ket{\psi_1}_n\ket{1} \right).
\end{equation}
Because we now have an extra factor of $2^p$, the polynomial $P$ in this case needs to approximate the function

\begin{equation}
	f(x) = \sqrt{\frac{Ae^{x^2 \cdot 2^p} - B}{C}},
\end{equation}
in order to generate the target state of Eq.~\eqref{eqn:general_exp_state}.

In order to handle negative values of $x$, we first need to pick a value $s \ge |x| \ge 0$ and then apply a binary addition circuit which performs $\ket{x}_n\ket{0}_m \rightarrow \ket{x}_n\ket{x+s}_m$, where $\ket{x+s}_m$ is again a fixed-point register with $m$ total digits and $p$ to the left of the binary point.
The unitary $\Usqrt$ of Eq.~\eqref{eqn:U_implementation_positive_scaling} is then applied with the second register as input to construct the state

\begin{equation}
	\label{eqn:U_implementation_negative_scaling}
	\Usqrt\ket{x+s}_m\ket{0}_{m+1} = \ket{x+s}_m\left(\sqrt{\frac{x+s}{2^p}}\ket{\psi_0}_m\ket{0} + \sqrt{1-\frac{x+s}{2^p}}\ket{\psi_1}_m\ket{1} \right),
\end{equation}
where the normalization factor $2^p$ arising from treating the fixed-point register $\ket{x+s}_m$ as a binary string in the comparator circuit, similarly handles the case $x \le -1.0$.
The polynomial $P$ which generates the target state of Eq.~\eqref{eqn:general_exp_state} needs to approximate the function

\begin{equation}
	\label{eqn:f_full_range}
	f(x) = \sqrt{\frac{Ae^{x^2 \cdot 2^p}e^{-s} - B}{C}}.
\end{equation}
Note that for both positive and negative values of $x$, the transformation in Eq.~\eqref{eqn:U_implementation_negative_scaling} guarantees that the input to the polynomial $P$ will only be the domain $x \in [0, 1]$.
This way, while only polynomials with definite parity (either even or odd) can be implemented using QSP, we can extend the function we approximate with QSP to the domain $x \in [-1, 1]$ using $f(-x) = f(x)$ or $f(-x)=-f(x)$, even if $f(x)$ does not have definite parity.
Additionally, the unitary $\Usqrt$ of Eq.~\eqref{eqn:U_implementation_negative_scaling} and the function Eq.~\eqref{eqn:f_full_range} allow us to construct the target state for any input $x$, provided that $f(x) \in [0, 1]$.

\section{Applications}
\label{sec:applications}
\subsection{European Call Option}
\label{sec:european_call}
In this section we illustrate how the QSP method described in the previous sections to prepare the state in Eq.~\eqref{eqn:general_exp_state} can be used in derivative pricing to compute payoffs of the form $f(S) = \max(0, S-K)$, by first considering a European call option.
Given a strike price $K$, the holder of a European call option contract receives the amount

\begin{equation}
	\label{eqn:european_call_payoff}
	f(S_T) = 		 \max(0, S_T - K),
\end{equation}
at a future date $T$ (the \emph{expiration date}), where $S_T$ is the market price of an underlying asset at $T$.
In order to determine the value of this contract today, the future price of the underlying asset is modeled as a stochastic variable with an appropriate probability distribution, and the price of the option is calculated by computing the expectation value of its \emph{payoff} in Eq.~\eqref{eqn:european_call_payoff}.

The re-parameterization method for derivative pricing described in Ref.~\cite{chakrabarti2021threshold}, implements the operator of Eq.~\eqref{eqn:derivative_path_loading} by loading a discretized normal distribution and then performing an affine transformation to the registers using quantum arithmetic to generate a superposition of possible asset log-returns $r_i$ at time $T$, each weighed by the probability of occurrence

\begin{equation}
	\label{eqn:prob_distribution_call}
	\sum_{i=0}^{N-1}\sqrt{p_i}\ket{r_i}_n,
\end{equation}
where $N=2^n$.
The price of the underlying asset for each possible log-return $r_i$ is given by $S_i = S_0e^{r_i}$, where $S_0$ is the price of the underlying asset today.
The payoff operator of Eq.~\eqref{eqn:derivative_payoff} is implemented by first calculating the price of the underlying asset from the log-return into another register

\begin{equation}
	\sum_{i=0}^{N-1}\sqrt{p_i}\ket{r_i}_n \rightarrow \sum_{i=0}^{N-1}\sqrt{p_i}\ket{r_i}_n\ket{S_i=S_0e^{r_i}},
\end{equation}
computing the payoff of Eq.~\eqref{eqn:european_call_payoff}

\begin{equation}
	\sum_{i=0}^{N-1}\sqrt{p_i}\ket{r_i}_n\ket{S_i} \rightarrow \sum_{i=0}^{N-1}\sqrt{p_i}\ket{r_i}_n\ket{S_i}\ket{f(S_i)},
\end{equation}
and finally encoding the payoff into the amplitude of an ancilla qubit

\begin{multline}
	\sum_{i=0}^{N-1}\sqrt{p_i}\ket{r_i}_n\ket{S_i}\ket{f(S_i)} \rightarrow \\ \sum_{i=0}^{N-1}\sqrt{p_i}\ket{r_i}_n\ket{S_i}\ket{f(S_i)} \left( \sqrt{\frac{\max(0, S_i- K)}{\fmax}} \ket{0} + \sqrt{\frac{1-\max(0, S_i - K)}{\fmax}}\ket{1} \right),
\end{multline}
where $\fmax=e^{\textrm{max}\{r_i\}} - K$ is the maximum value of the payoff possible by the discretization chosen in Eq.~\eqref{eqn:prob_distribution_call}.
The probability of measuring $\ket{0}$ in the last qubit is

\begin{equation}
	\label{eqn:reparam_price}
	\mathbb{P}[\ket{0}] = \sum_{i=0}^{N-1}p_i\frac{\max(0, S_i - K)}{\fmax},
\end{equation}
which is the expectation value of the option's payoff, normalized by $\fmax$.
Thus, applying amplitude estimation to the last qubit allows us to estimate the price of the option.

\begin{figure}[t]
\centering
\includegraphics[width=0.7\linewidth]{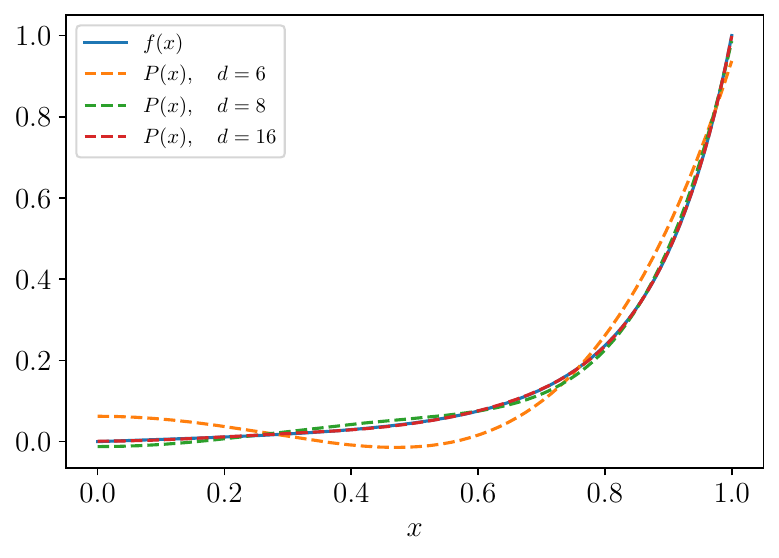}
\caption{The function $f(x)$ of Eq.~\eqref{eqn:f_call_option} with $S_0=K=100$, $p=3$, $s=0$ and $\fmax=S_0e^{2^p} - K$ used to compute the payoff of a European call option, and approximating polynomials $P(x)$ of degree $d \in [6,8,16]$, which can be applied to a quantum system using the QSP unitary of Eq.~\eqref{eqn:qsp_general_transformation}. The polynomial approximations are generated using the method described in Appendix~\ref{app:qsp_approximation}. }%
\label{fig:call_option_func_approx}
\end{figure}

We now describe how QSP and the methods developed in the previous sections can be used to perform the same computation.
Let the $\ket{r_i}_n$ register of Eq.~\eqref{eqn:prob_distribution_call} represent each log-return in fixed-point representation and let $s=\ln(S_0/K)$.
Apply a circuit which performs the comparison $\ket{r_i}_n\ket{0} \rightarrow \ket{r_i}_n \ket{r_i \le s}$ to Eq.~\eqref{eqn:prob_distribution_call} to get

\begin{equation}
	\sum_{\{i | r_i > s \}}\sqrt{p_i}\ket{r_i}_n\ket{0} + \sum_{\{i | r_i \le s \}}\sqrt{p_i}\ket{r_i}_n\ket{1}.
\end{equation}
Add an $m$-qubit register and compute $\ket{r_i}_n\ket{0}_m \rightarrow \ket{r_i}_n\ket{r_i+|s|}_m$, where $m$ is large enough to hold the maximum value of $r_i+|s|$ in fixed-point representation, to get

\begin{equation}
	\sum_{\{i | r_i > s \}}\sqrt{p_i}\ket{r_i}_n\ket{r_i+|s|}_m\ket{0} + \cdots,
\end{equation}
where we ignore terms with the last qubit being in the $\ket{1}$ state.
Note that $0 \le r_i+|s| < 2^p$ if the register $\ket{r_i+|s|}_m$ has $p$ digits to the left of the binary point.
The application of the $\Usqrt$ operator of Eq.~\eqref{eqn:U_implementation_negative_scaling} to the $\ket{r_i+|s|}_m$ register and a new qubit register $\ket{0}_{m+1}$ gives

\begin{equation}
	\sum_{\{i | r_i > s \}}\sqrt{p_i}\ket{r_i}_n\ket{r_i+|s|}_m\left(\sqrt{\frac{r_i+|s|}{2^p}}\ket{\psi_0}_m\ket{0} + \sqrt{1-\frac{r_i+|s|}{2^p}}\ket{\psi_1}_m\ket{1}\right) \ket{0},
\end{equation}
for normalized quantum states $\ket{\psi_0}$ and $\ket{\psi_1}$.
Employing the QSP unitary $U_{+}^{\vec{\phi}}$ of Eq.~\eqref{eqn:qsp_general_transformation} with phases $\vec{\phi}=(\phi_1, \phi_2, \cdots, \phi_d)$ chosen such that the polynomial $P$ approximates the function

\begin{equation}
	\label{eqn:f_call_option}
	f(x) = \sqrt{\frac{S_0e^{x^2\cdot 2^p}e^{-|s|} - K}{\fmax}}
\end{equation}
produces

\begin{equation}
	\sum_{\{i | r_i > s \}}\sqrt{p_i}\ket{r_i}_n\ket{r_i+|s|}_m\left(\sqrt{\frac{S_0e^{r_i} - K}{\fmax}}\ket{\psi'_0}_m\ket{G} \right),
\end{equation}
with $\ket{G} = \ket{0}_3$ for odd $d$, and

\begin{equation}
	\sum_{\{i | r_i > s \}}\sqrt{p_i}\ket{r_i}_n\ket{r_i+|s|}_m\left(\sqrt{\frac{S_0e^{r_i} - K}{\fmax}}\ket{G}\right),
\end{equation}
with $\ket{G} = \ket{0}_{m+3}$ for even $d$, where we have ignored terms with any qubit in the last register being in the $\ket{1}$ state.
Applying amplitude estimation to estimate the probability of measuring the $\ket{G}$ state gives us

\begin{equation}
	\mathbb{P}[\ket{G}] = \sum_{\{i | r_i > s \}}p_i\frac{S_0e^{r_i} - K}{\fmax},
\end{equation}
which is the same as the value we estimate with the standard re-parameterization method in Eq.~\eqref{eqn:reparam_price}.
In this case however, we do not explicitly compute the values $\ket{S_i = S_0e^{r_i}}$ or $\ket{f(S_i)}$ using quantum arithmetic, but rather implicitly calculate them by applying amplitude transformations using QSP.
In Fig.~\ref{fig:call_option_func_approx}, we see plot the function of Eq.~\eqref{eqn:f_call_option} with $S_0=K=100$, $p=3$, $s=0$ and $\fmax=S_0e^{2^p} - K$ along with polynomial approximations of varying degrees constructed using the optimization-based method described in Appendix~\ref{app:qsp_approximation}.

We note that Ref.~\cite{mcardle2022quantum} discussed the possibility of applying QSP to compute common derivative payoffs such as that of Eq.~\eqref{eqn:european_call_payoff}.
The authors pointed out that polynomial approximations to functions of the form $f(x)=\sqrt{x}$ suffer due to the fact that the function is not differentiable at the origin.
By including the exponential in the function we are approximating, not only do we pass more computation to the QSP transformation, but we also address this issue since Eq.~\eqref{eqn:f_call_option} is differentiable in the entire domain of applicability.

\subsection{Autocallable}
\label{sec:autocallable}
For a single underlying asset, an \emph{autocallable} derivative contract consists of

\begin{enumerate}
    \item A set of $k$ binary payoffs $\{K_i, t_i, f_i \}$, such that if the asset returns at time $t_i$ exceed $K_i$, the holder of the contract receives a return $f_i$ and the contract terminates. The asset's return at time $t$ is defined as $S_t/S_0$, where $S_t$ is the value of the asset at time $t$ and $S_0$ the value of the asset today.
    \item If none of the binary payoffs have been triggered, and the asset returns have crossed a barrier value $B$ at any time before the contract's expiration, the holder receives a return given by $1-\max(0, K_T - f_T)$, where $K_T \in [0,1]$ is a pre-determined \emph{strike} return and $f_T$ is the asset's return at expiration date $T$.
\end{enumerate}
Because an autocallable is typically defined in terms of an asset's performance, it also includes a (dollar) notional value $V$ on which the payoff returns are evaluated to determine a dollar amount to be exchanged.
For example, if the contract's notional value is $\$1$M, and a binary payoff with $f_i = 1.5$ is triggered, the holder of the contract receives $\$1.5$M in total.
We now describe how this contract can be priced by employing QSP at the appropriate place.

Starting again from the re-parameterization method from \cite{chakrabarti2021threshold}, we first initialize a superposition of probability-weighted possible paths for the asset's log-returns at each timestep

\begin{equation}
	\sum_{r}\sqrt{p_{r}}\ket{r} \quad \textrm{ with } \quad \ket{r} = \bigotimes_{t=1}^{T}\ket{r_{t}},
\end{equation}
where the asset's total return at timestep $t$ is given by $e^{r_t}$.
We add a $k$-qubit register $\ket{s}$ and on each qubit $s_i$ we calculate whether the $k$-th binary payoff is triggered, by checking whether the condition $r_{t_i} > \ln K_{t_i}$ is satisfied and no previous binary payoffs have been triggered

\begin{equation}
	\label{eqn:binary_payoff_indicators}
	\ket{s} = \bigotimes_{i=1}^k \Ket{s_i = \left((r_{t_i} > \ln K_{t_i}) \cdot (\neg s_{i-1}) \right)},
\end{equation}
where the $\neg$ symbol denotes logical negation and $s_{0}=0$.
Each binary variable $s_i$ will thus be $1$ if the asset return at time $t_i$ exceeds the threshold amount $K_i$ and none of the previous payoffs have been triggered, hence indicating whether the payoff at $t_i$ should be payed.
To determine whether the payoff from the second autocallable condition is triggered, we must check if the barrier value $B$ has been crossed at any timestep of the simulation $t$.
This is achieved by first adding a $(T+1)$-qubit register $\ket{b}$ and calculating $b_t = r_t \le \ln B $ in each qubit for $t \in [1,T]$, and then computing

\begin{equation}
	\label{eqn:knock_indicator}
b_{T+1} = \left(\neg\prod_{t=1}^{T}b_i \right) \cdot \prod_{i=1}^{k}\left(\neg s_i\right),
\end{equation}
such that $b_{T+1} = 1$ if $r_t > B$ for at least one value of $t$, and every qubit in $\ket{s}$ is zero.
Therefore, $b_{T+1}=1$ indicates that the barrier $B$ has been crossed at least once before the contract's expiration, and no binary payoffs have been triggered, meaning the second condition of the autocallable payoff is satisfied.
Note that all these logical operations above can be computed using just CNOT and Toffoli gates.

Un-computing and dropping the first $T$ qubits of the $\ket{b}$ register, we arrive at the state

\begin{equation}
	\sum_{r}\sqrt{p_{r}}\ket{r} \bigotimes_{i=1}^{k}\ket{s_i}\ket{b}.
\end{equation}
The $\mathcal{F}$ operator needed to encode the autocallable contract's payoff into the amplitude of an ancilla qubit as outlined in Eq.~\eqref{eqn:derivative_payoff}, should construct the state (up to normalization)

\begin{equation}
    \label{eqn:autocallable_payoff_operator}
	\sum_{r}\sqrt{p_{r}}\ket{r} \bigotimes_{i=1}^{k}\ket{s_i}\ket{b}   \begin{cases}
					\left(\sqrt{f_i}\ket{0} + \sqrt{1-f_i}\ket{1} \right), & \text{if $s_i=1$} \\
                    \left(\sqrt{1-\max(0, K_T - f_T)}\ket{0} + \sqrt{\max(0, K_T - f_T)}\ket{1} \right), & \text{if $b=1$} \\
					\ket{1}, & \text{otherwise}
	\end{cases}
\end{equation}
where $f_T = e^{r_T}$ is the asset's return at the final timestep of the computation.
The first payoff clause can be implemented using controlled Ry rotations since the values $f_i$ are known in advance of the computation.
For the second clause, we first apply a comparator circuit with an ancilla qubit to calculate $\ket{r_T < \ln K_T}$, and then controlled on this qubit, we need to prepare the state

\begin{equation}
	\label{eqn:autocallable_qsp_transformation}
    \ket{r_T}_n\ket{0}_{n+q} \rightarrow \ket{r_T}_n\left(\sqrt{1-\left(K_T - e^{r_T}\right)}\ket{\psi_0}_n\ket{0}_q +  \sqrt{\left(K_T - e^{r_T}\right)}\ket{\psi_1}_n\ket{0_{\perp}}_q\right),
\end{equation}
where $\ket{\psi_0}_n$, $\ket{\psi_1}_n$ are arbitrary normalized states.
Note that we have already imposed the condition $1-\left(K_T - e^{r_T}\right) \in [0,1]$, so the state is appropriately normalized.
Since this transformation has the same form as Eq.~\eqref{eqn:general_exp_state}, we can approximately prepare it using the QSP method described in Sections \ref{sec:main_idea} and \ref{sec:implementation}, similarly to the European call option described in Sec.~\ref{sec:european_call}.
Assuming $\ket{r_T}_n$ is an $n$-qubit fixed-point representation of $r_T$ with $p$ digits to the left of the binary point and spans over both positive and negative values, we employ the unitary of Eq.~\eqref{eqn:U_implementation_negative_scaling} with an appropriately chosen value $s \ge \max(|r_T|) \ge 0$, to first generate the state

\begin{equation}
	\label{eqn:U_implementation_autocallable}
	\Usqrt\ket{r_T+s}_m\ket{0}_{m+1} = \ket{r_T+s}_m\left(\sqrt{\frac{r_T+s}{2^p}}\ket{\psi_0}_m\ket{0} + \sqrt{1-\frac{r_T+s}{2^p}}\ket{\psi_1}_m\ket{0_{\perp}} \right).
\end{equation}
Subsequently, the polynomial transformation $P$ we apply with QSP in order to generate Eq.~\eqref{eqn:autocallable_qsp_transformation} approximates the function

\begin{equation}
	\label{eqn:f_autocallable}
	f(x) = \sqrt{1- \left(K_T-e^{x^2 \cdot 2^p}e^{-s} \right)}.
\end{equation}

While we have described an autocallable contract defined only on a single underlying asset, the same approach can be extended to multi-asset cases such as \emph{BestOf} and \emph{WorstOf} variants.
In those cases, we generate stochastic paths for all underlying assets and apply a circuit to compare the relative performance of all assets, identifying the best/worst performing asset.
The payoff is then implemented similarly to the single-asset case following Eq.~\eqref{eqn:autocallable_payoff_operator}, based on the returns of the identified asset.

\section{Resource Estimates for Quantum Advantage}
\label{sec:resources}
We evaluate the efficiency of employing QSP in the evaluation of derivative payoffs by estimating the resources needed to price an autocallable derivative contract using the method described in Sec.~\ref{sec:autocallable}.
As a benchmark we use the resource estimates from Ref.~\cite{qsharpblog}, based on the reparameterization method introduced in Ref.~\cite{chakrabarti2021threshold} where coherent quantum arithmetic is used to compute the derivative payoff.

The goal of the reparameterization method is to prepare a probability distribution across $a$ stochastic underlying assets and $T$ timesteps under the assumption that asset prices follow geometric Brownian motion.
In this setting, asset prices are described by a multivariate lognormal distribution, or equivalently, the asset log-returns are distributed normally.
The reparameterization method prepares the distribution over log-returns by first loading independent standard normal distributions $\mathcal{N}(0,1)$ in $a \cdot T$ quantum registers, and then applying quantum arithmetic across the quantum registers to generate the multivariate normal distribution $\mathcal{N(\bm{\mu}, \mathbf{\Sigma})}$ representing the distribution of log-returns at each timestep.
The parameters $(\bm{\mu}$, $\mathbf{\Sigma})$ are typically determined by historical market values and characterize the stochastic process of each asset.
The price of each asset at each timestep is then calculated by exponentiating the resulting quantum registers, and the payoff is computed into another register based on the asset prices, before encoding the result into the amplitude of an ancilla qubit.

\begin{figure}[t]
\centering
\subfloat[]{\label{fig:reparam_circuit}\includegraphics[width=.42\linewidth]{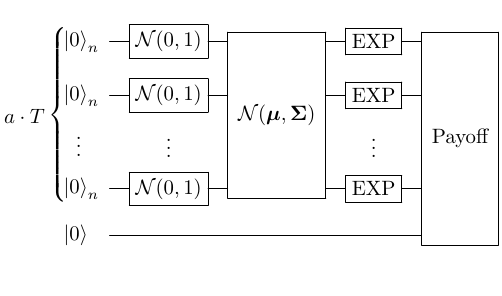} }%
\subfloat[]{\label{fig:reparam_circuit_qsp}\includegraphics[width=.5\linewidth]{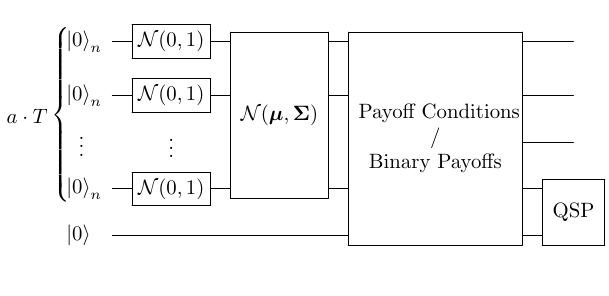} }%
\caption{Schematic representation of the quantum circuits used to price the autocallable derivative contract using \textbf{(a)} the standard reparameterization method of Ref.~\cite{chakrabarti2021threshold} and \textbf{(b)} the QSP approach described in Sec.~\ref{sec:autocallable}. The two approaches employ the same recipe to prepare the multivariate normal distribution $\mathcal{N(\bm{\mu}, \mathbf{\Sigma})}$ for each timestep $t \in [1,T]$ across $d$ underlying assets. The application of QSP in the evaluation of the payoff via Eq.~\eqref{eqn:autocallable_payoff_operator} removes the necessity of directly computing exponentials of quantum registers, a major contributor to the quantum resources required to implement the circuit \cite{chakrabarti2021threshold}. The quantum circuits depict only the main quantum registers involved in the calculation and not ancillary registers that may be required in the subcomponents. }%
\label{fig:reparam_circuits}
\end{figure}

In the QSP approach outlined in Sec.~\ref{sec:autocallable}, we again employ the reparameterization method to prepare the distribution $\mathcal{N(\bm{\mu}, \mathbf{\Sigma})}$ over the assets' log-returns.
We subsequently evaluate the payoff conditions of Eq.~\eqref{eqn:binary_payoff_indicators} and Eq.~\eqref{eqn:knock_indicator} using ancilla qubits, and encode the binary payoffs to the payoff ancilla qubit controlled on the corresponding payoff condition qubits.
Note that the above does not require direct knowledge of asset prices, and can all be performed using only the log-returns, thus we do not need to calculate exponentials of values stored in binary representation in quantum registers.
At the final timestep, we use QSP to implement the second clause of Eq.~\eqref{eqn:autocallable_payoff_operator} which encodes the payoff directly in the amplitude of the payoff qubit, and as such we eliminate all instances of quantum register exponentiation in the circuit.
A schematic representation of the quantum circuits implementing the standard reparameterization method of Ref.~\cite{chakrabarti2021threshold} and the corresponding QSP approach outlined in Sec.~\ref{sec:autocallable} in order to price an autocallable contract is shown In Fig.~\ref{fig:reparam_circuits}.

The resource estimates in Ref.~\cite{chakrabarti2021threshold, qsharpblog} for quantum advantage in pricing an autocallable contract are computed for an instance with $a=3$ assets and $T=20$ timesteps, and the overall target approximation error in the estimate is set to $\epsilon = 2 \times 10^{-3}$.
The corresponding accuracy with which the payoff must be calculated and encoded into the amplitude of an ancilla qubit is $\epsilon_p = 10^{-3}$, meaning that the unitary $\mathcal{F}$ of Eq.~\eqref{eqn:derivative_payoff} must be implemented such that the amplitude of the $\ket{0}$ is accurate to within $\epsilon_p$.
In the standard reparameterization circuit of Fig.~\ref{fig:reparam_circuit}, this target accuracy dictates the resources required to compute the exponentials, the quantum arithmetic in the computation of the payoff and the final rotation into an ancilla qubit using controlled-$Ry$ gates.
By virtue of Eq.~\eqref{eqn:autocallable_qsp_transformation}, the QSP process encompasses the computation of the exponential and the payoff simultaneously, as well as the amplitude encoding once we have access to the unitary of Eq.~\eqref{eqn:U_implementation_autocallable} which encodes the value $\sqrt{r_T}$ in the amplitude of an ancilla qubit.
The corresponding payoff resources of the QSP circuit in Fig.~\ref{fig:reparam_circuit_qsp} are determined by the accuracy of the transformation of Eq.~\eqref{eqn:autocallable_qsp_transformation}, which in turn depends on the degree $d$ of the polynomial $P(x)$ required to approximate the function $f(x)$ of Eq.~\eqref{eqn:f_autocallable} such that $\max|f(x) - P(x)| \le \epsilon_p = 10^{-3}$.

\begin{figure}[t]
\centering
\subfloat[]{\label{fig:autocallable_approximation}\includegraphics[width=.47\linewidth]{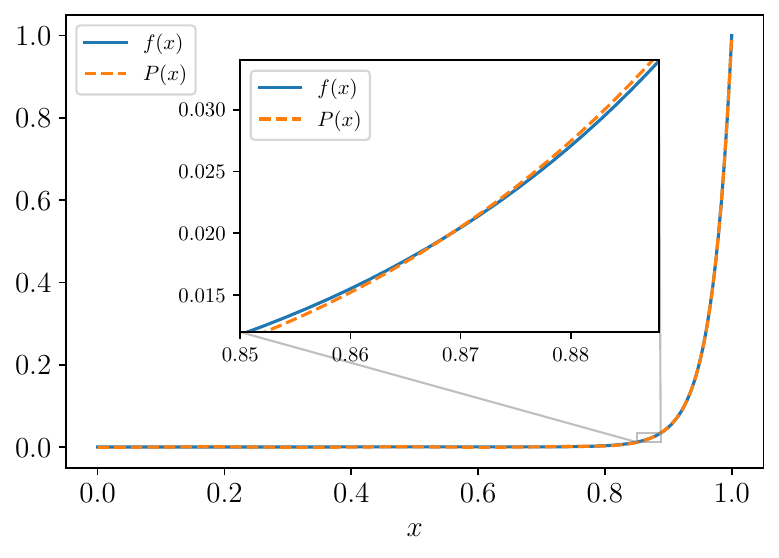} }%
\subfloat[]{\label{fig:autocallable_approximation_error}\includegraphics[width=.51\linewidth]{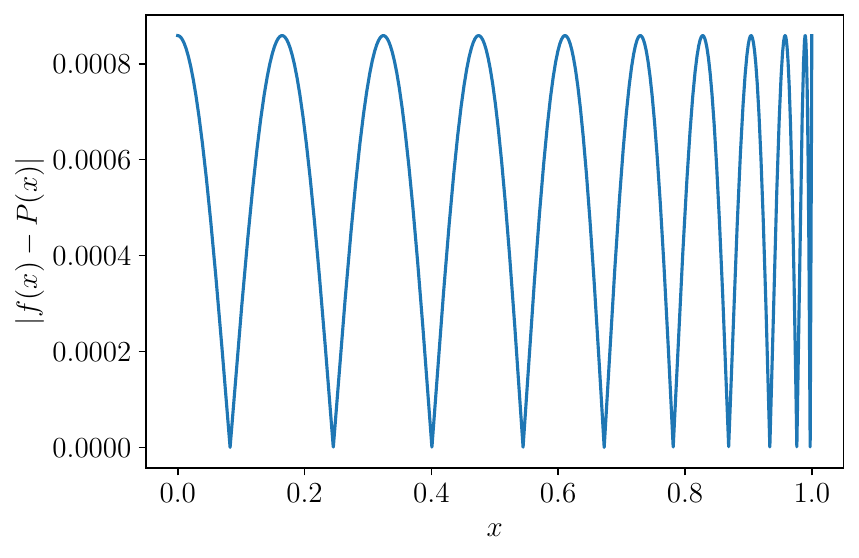} }%
\caption{\textbf{(a)} The function $f(x)$ of Eq.~\eqref{eqn:f_autocallable} with $p=5$, $s=2^p$ and $K_T=1$ used to compute the payoff of an autocallable contract, and a $d=20$-degree approximating Chebyshev polynomial $P(x)$ generated using the method described in Appendix~\ref{app:qsp_approximation}. The QSP transformation of Eq.~\eqref{eqn:qsp_general_transformation} based on the unitary $\Usqrt$ of Eq.~\eqref{eqn:U_implementation_autocallable} can then be employed to approximately create the state in Eq.~\eqref{eqn:autocallable_qsp_transformation}, encoding the autocallable payoff directly in a quantum amplitude. \textbf{(b)} The absolute error between $f(x)$ and the approximation $P(x)$ satisfies the target accuracy $\max|f(x) - P(x)| \le 10^{-3}$.}%
\label{fig:autocallable_approx}
\end{figure}

We extend the Q\# \cite{QsSpec2020} implementation from Ref.~\cite{qsharpblog} to construct the quantum circuit of Fig.~\ref{fig:reparam_circuit_qsp} and estimate the corresponding quantum resources when QSP is used to compute the exponentials and derivative payoff, for both $\Usin$ (Eq.~\eqref{eqn:sine_amplitude_oracle}) and $\Usqrt$ (Eq.~\eqref{eqn:sqrt_amplitude_oracle}) as the underlying unitary.
In the original implementation, the parameters of the quantum circuit up to and including the block labeled $\mathcal{N(\bm{\mu}, \mathbf{\Sigma})}$ were chosen to satisfy the target accuracy requirements and are thus kept the same.
As such, the register $\ket{r_T}$ uses the fixed-point representation of Eq.~\eqref{eqn:fixed_point_repr} with $(n, p)=(15, 5)$ such that $r_T \in [-R, R]$ with $R=2^5$.
Choosing $s=R$ in Eq.~\eqref{eqn:U_implementation_autocallable} makes sure the encoded amplitudes are positive, and the $\Usqrt$ operator is constructed using the method described in Sec.~\ref{sec:implementation} with a ripple-carry comparator circuit \cite{draper2006logarithmic}.
Using the polynomial approximation construction described in Appendix~\ref{app:qsp_approximation}, we numerically determine that a polynomial degree $d=20$ suffices to generate a polynomial $P(x)$ which satisfies $\max|f(x) - P(x)| \le 10^{-3}$ for all values of $K_T \in [0,1]$.
With these parameters, in Fig.~\ref{fig:autocallable_approx}, we plot the function $f(x)$ and the generated polynomial approximation $P(x)$, as well as the corresponding approximation error $|f(x) - P(x)|$ for $K_T=1$.

The resource estimation from Ref.~\cite{chakrabarti2021threshold} was carried out with the assumption that piecewise polynomial approximation \cite{hner2018optimizing} is used to compute the exponentials required in the circuit.
In Ref.~\cite{qsharpblog}, the quantum circuits were explicitly constructed, increasing the accuracy of the estimates, and the exponentials were computed using a lookup table.
The lookup table is parameterized by the number of swap bits, offering tradeoffs between the circuit depth/width and the number of qubits required.
The quantum resources reported are that of the $\mathcal{Q}$ operator driving amplitude estimation, where $\mathcal{Q} = \mathcal{A}S_0\mathcal{A}^{\dagger}S_1$, for appropriate reflection operators $S_0$ and $S_1$ \cite{brassard2002quantum}, and the $\mathcal{A}$ operator of Eq.~\eqref{eqn:A_operator}.
In Table~\ref{tab:resources}, we show the $T$-depth, $T$-count and number of logical qubits required to implement the $\mathcal{Q}$ operator with \textbf{(a)} the approach from Ref.~\cite{qsharpblog} where a lookup table with one swap qubit is used to compute the exponentials, while arithmetic is used to calculate the payoff, and \textbf{(b)} the approach developed in this work where the exponentials and payoff are computed using QSP either with the unitary of Eq.~\eqref{eqn:sine_amplitude_oracle} or the unitary of Eq.~\eqref{eqn:sqrt_amplitude_oracle}.
We notice that the QSP method based on either unitary significantly outperforms the Arithmetic approach, but because the $\Usin$-based QSP includes $d \cdot n$ total controlled-Ry rotations ($n$ controlled-Ry rotations for each invocation of $\Usin$ and $d$ total invocations of $\Usin$ and $\Usin^{\dagger}$ in the QSP circuit), the total $T$-depth is significantly larger compared to using $\Usqrt$.
In the Arithmetic column we use the results from Ref.~\cite{qsharpblog} with one swap qubit for reference as it gives the most balanced resources for the metrics considered, but note that the $\Usqrt$-based QSP method  outperforms the arithmetic method also in the case where five or ten swap qubits are used for the lookup table across all three metrics.

The difference in T-depth between the two QSP approaches can be understood by considering the Clifford+T decomposition of both $\Usin$ and $\Usqrt$.
The unitary $\Usqrt$ can be implemented with a Toffoli-depth of $2\lfloor \log_2(n-1)\rfloor + 5$ using a ripple-carry comparator for an $n$-qubit register \cite{draper2006logarithmic}, which we can also regard as the equivalent T-depth if we use an ancilla qubit for the T-gate decomposition of each Toffoli \cite{Selinger_2013}.
Thus, $\Usqrt$ can be constructed with a T-depth of $11$ for the $n=15$ case we consider here.
On the other hand, the $\Usin$ unitary requires Ry rotations, each of which can be performed to precision $\epsilon_R$ with a circuit of T-depth $3\log_2(1/\epsilon_R)$ \cite{ross_2016}.
Since there are $d \cdot n$ total Ry rotations in the QSP circuit, if we want the total error from all rotations to be within our total target payoff error $\epsilon_t \le \epsilon_p = 10^{-3}$, each Ry rotation needs to be performed with precision $\epsilon_R = \epsilon_t/(d \cdot n) \sim 3 \times 10^{-6}$.
Hence, each $\Usin$ unitary requires a T-depth of $15 \cdot 3\log_2(1/\epsilon_R) \sim 818$, about $74$ times larger than that of $\Usqrt$.

Using the Iterative Quantum Amplitude Estimation method \cite{grinko2021iterative} to compute the total resources needed for the end-to-end process to estimate the target amplitude to within $\epsilon = 10^{-3}$ at confidence level $1-\alpha=0.68$ \cite{chakrabarti2021threshold}, we find that the $\Usqrt$-QSP method requires $4.7$k logical qubits, a total T-depth of $4.5 \times 10^7$ and T-count of $2.4 \times 10^9$.
With the assumption that this derivative contract can be priced classically in one second \cite{chakrabarti2021threshold}, a quantum processor would need to execute T-gates at a rate of $45$MHz to match the classical performance if QSP with $\Usqrt$ is used, compared to $207$MHz with the arithmetic method and $430$MHz using QSP and $\Usin$.

\begin{table*}
\centering
	\begin{tabularx}{\textwidth}{@{}l *5{>{\centering\arraybackslash}X}@{}}\toprule
	\textbf{Resource} & \textbf{Arithmetic} \cite{qsharpblog}  &\multicolumn{2}{c}{\textbf{QSP (this work)}} & \textbf{Improvement}
	\\\cmidrule(lr){3-4}
		&  &$\Usin$ [Eq.~\eqref{eqn:sine_amplitude_oracle}] & $\Usqrt$ [Eq.~\eqref{eqn:sqrt_amplitude_oracle}]  & \\\midrule
		\textbf{$T$-depth} & $36$k & $75$k & $7.8$k  & 4.7x\\
		\textbf{$T$-count} & $6.6$M & $605$k & $414$k  & $16$x\\
		\textbf{\# Qubits}  & $19.2$k & $4.7$k & $4.7$k & $4.1$x\\\bottomrule
	\end{tabularx}
	\caption{Quantum resources required to implement the $\mathcal{Q}$ operator driving Amplitude Estimation to price the autocallable contract from Sec.~\ref{sec:resources} with $a=3$ assets and $T=20$ timesteps. The Arithmetic column refers to the results from Ref.~\cite{qsharpblog} where the exponentials are computed using a lookup table with one swap qubit, and the payoff using quantum arithmetic \cite{chakrabarti2021threshold}. In the QSP column, the corresponding resources are shown when QSP is employed to compute both the exponentials and the payoff as described in Sec.~\ref{sec:autocallable}, based on either Eq.~\eqref{eqn:sine_amplitude_oracle} or Eq.~\eqref{eqn:sqrt_amplitude_oracle} as the underlying unitary. In the last column we highlight the improvement of $\Usqrt$-QSP for each resource compared to the Arithmetic column. For all cases, the resources are calculated such that all encoded amplitudes are accurate to within $\epsilon_p \le 10^{-3}$.}
    \label{tab:resources}
\end{table*}

\section{Discussion}
\label{sec:discussion}
It has been previously highlighted that quantum arithmetic is a major bottleneck in pricing financial derivatives on quantum computers \cite{chakrabarti2021threshold, qsharpblog}, and the QSP-based method we have introduced to construct derivative payoffs significantly reduces the quantum resources required for quantum advantage compared to previous methods, in both circuit width and depth.
This approach seems more natural in the following sense: When an application requires the encoding of a function of a variable into a quantum amplitude, it should be more efficient to directly transform the amplitude, rather first applying a digital transformation through arithmetic on computational basis states and subsequently encoding the result into the amplitude.
While QSP restricts the domain and range of transformations to the interval $[0,1]$, we observe that in cases like the one studied in this manuscript it is still advantageous compared to utilizing additional quantum resources to increase the range/precision of digital operations.
We are therefore hopeful that similar techniques can be explored in other use cases, potentially leading to additional improvements to the method.
While the topic we are addressing in this manuscript is quite different from that of Ref.~\cite{mcardle2022quantum}, the unitary we introduced in Eq.~\eqref{eqn:sqrt_amplitude_oracle} can also be employed for the state preparation method introduced in that work, potentially further reducing the resources in that case as well.
Additionally, the availability of open source implementations for QSP phase factor calculations \cite{qsppack, pyqsp} enables the exploration of QSP for specific applications more accessible, without requiring deep expertise in the topic.

One limitation of the QSP method in its current form is that it can only provide approximations to functions of a single variable.
While for the applications considered in this work this is enough to provide an advantage, the ability to handle multivariate functions would greatly increase its applicability.
For example, multivariable QSP (M-QSP) could potentially reduce the quantum resources in derivative pricing further by replacing the quantum arithmetic in the circuit component labeled $\mathcal{N(\bm{\mu}, \mathbf{\Sigma})}$ in Fig.~\ref{fig:reparam_circuits}.
Moreover, it would enable the computation of more complicated derivative payoffs which are non-trivial functions of multiple stochastic variables.
Some properties and aspects of M-QSP have been discussed in Ref.~\cite{Rossi_2022}, and additional research on this topic would have potentially far-reaching practical consequences for quantum applications.
We believe that this is a very worthwhile topic for further research.

\section{Acknowledgments}
We thank Rajiv Krishnakumar, Dave Clader and Farrokh Labib for useful discussions on derivative pricing and QSP, and Lin Lin for pointing us to QSPPACK to numerically generate polynomial approximations and phase factors used in QSP.

\bibliographystyle{apsrev4-1_custom}
\bibliography{qsp_payoff}

\appendix
\section{Construction of Polynomial Approximations for QSP}
\label{app:qsp_approximation}
In order to approximately apply a singular value transformation given by a function $f(x)$ using QSP, we first need to find a polynomial which approximates it.
The polynomial transformations possible through the QSP framework must have definite parity (even or odd), but because we only aim to transform real positive amplitudes created by the unitary of Eq.~\eqref{eqn:U_implementation_negative_scaling}, we restrict our attention to the interval $[0,1]$ due to symmetry.

We employ the optimization-based method described in Ref.~\cite{dong2022ground} which generates near-optimal approximations without relying on any analytic computation.
This method constructs a linear combination of even Chebyshev polynomials with some unknown coefficients $\{c_k\}$

\begin{equation}
	P(x)=\sum_{k=0}^{d/2}c_kT_{2k}(x),
\end{equation}
and aims to find the coefficients $\{c_k\}$ which give the best approximation to $f(x)$.
This process is formulated as a discrete optimization problem by discretizing the interval $[0, 1]$ using $M$ grid points generated by the roots of Chebyshev polynomials $\left\{x_j = -\cos\frac{j\pi}{M-1}\right\}_{j=0}^{M-1}$ and solving the following minimax optimization problem for the coefficients $\{c_k\}$

\begin{eqnarray}
	\label{eqn:threshold_optimization}
	&\underset{\{c_k\}}{\min} \quad \left\{ \underset{x_j \in [0, 1]}{\max} |f(x_j)-P(x_j)|\right\} \nonumber \\
	& \textrm{s.t.} \quad P(x_j)=\sum_k c_kA_{jk}, \quad |P(x_j)| \le c, \quad \textrm{for } j=0, 1, \ldots, M-1,
\end{eqnarray}
where $c$ is chosen close to $1$ but preferably slightly smaller to avoid overshooting, and $A_{jk}$ is a matrix of coefficients defined as $A_{jk}=T_{2k}(x_j)$ for $k=0, 1, \ldots, d/2$.

\end{document}